%% file: main.tex
\documentclass[sigconf]{acmart}

\input{header}

\input{nomenclature}

\AtBeginDocument{%
  \providecommand\BibTeX{{%
    \normalfont B\kern-0.5em{\scshape i\kern-0.25em b}\kern-0.8em\TeX}}}

\setcopyright{acmcopyright}
\copyrightyear{2018}
\acmYear{2018}
\acmDOI{10.1145/1122445.1122456}

\acmConference[Woodstock '18]{Woodstock '18: ACM Symposium on Neural
  Gaze Detection}{June 03--05, 2018}{Woodstock, NY}
\acmBooktitle{Woodstock '18: ACM Symposium on Neural Gaze Detection,
  June 03--05, 2018, Woodstock, NY}
\acmPrice{15.00}
\acmISBN{978-1-4503-XXXX-X/18/06}

\begin{document}

\title{ Demonstration of \sys: Interactive Visualization Interface Generation for SQL Analysis in Notebook}

\author{Jeffrey Tao}
\email{jat2164@columbia.edu}
\affiliation{
  \institution{Columbia University}
  \city{New York}
  \state{NY}
  \country{USA}
}
\author{Yiru Chen}
\email{yiru.chen@columbia.edu}
\affiliation{
  \institution{Columbia University}
  \city{New York}
  \state{NY}
  \country{USA}
}
\author{Eugene Wu}
\email{ewu@cs.columbia.edu}
\affiliation{
  \institution{Columbia University}
  \city{New York}
  \state{NY}
  \country{USA}
}

\begin{abstract}
We demonstrate PI2, the first notebook extension that can auto- matically generate interactive visualization interfaces during SQL-based analyses.

\end{abstract}

\maketitle

\input{introduction.tex}

\input{overview.tex}

\input{demontration.tex}

\bibliographystyle{ACM-Reference-Format}
\bibliography{ref}

\end{document}

%% file: header.tex
\usepackage{amsmath}
\usepackage{graphicx}
\usepackage{balance}  %
\usepackage{adjustbox}
\usepackage{footmisc}
\usepackage{soul}
\usepackage{hyperref}
\usepackage{enumitem}
\usepackage{multirow}
\usepackage{listings}
\usepackage{cleveref}
\usepackage{array}
\usepackage{tabularx}
\usepackage{booktabs} %
\usepackage{xspace}
\usepackage{xcolor}
\usepackage{microtype}
\usepackage{subcaption}
\usepackage{balance}
\usepackage{mathtools}
\usepackage{scalerel}
\usepackage{adjustbox}
\usepackage{float}
\usepackage[noend]{algpseudocode}
\usepackage{algorithmicx,algorithm}
\usepackage{bbding}

\definecolor{darkgreen}{rgb}{0.15,0.55,0.15}
\definecolor{darkblue}{rgb}{0.1,0.1,0.5}
\definecolor{orange}{rgb}{1, 0.64, 0}
\definecolor{blue}{rgb}{0.01,0.40,.8}
\definecolor{darkgreen}{rgb}{0.15,0.55,0.15}
\definecolor{mred}{rgb}{.80,.12,.30}
\definecolor{grey}{rgb}{0.5,0.5,0.5}
\definecolor{Purple}{rgb}{.75,0,.85}
\definecolor{light-gray}{gray}{0.95}
\definecolor{mid-gray}{gray}{0.85}
\definecolor{darkred}{rgb}{0.7,0.25,0.25}
\definecolor{rose}{rgb}{1.0, 0.01, 0.24}
\definecolor{lightyellow}{HTML}{FFD700}
\definecolor{lgreen}{HTML}{C2E9D7}
\definecolor{aqua}{rgb}{0.03, 0.91, 0.87}
\definecolor{rose}{rgb}{0.9, 0.13, 0.13}
\newcommand{\red}[1]{\textcolor{red}{#1}}

\newcommand{\gray}[1]{\textcolor{grey}{#1}}

\newcommand{\eat}[1]{}

\newcommand{\ititle}[1]{\vspace{2pt}\noindent\emph{#1}}
\newcommand{\stitle}[1]{\smallskip\noindent\textbf{#1}}

\newlength{\listingindent}                %
\usepackage{tablefootnote}

\usepackage[LGRgreek]{mathastext}

%% file: nomenclature.tex
\newtheorem{example}{Example}

\newcommand{\sys}{\textsc{PI2}\xspace}
\newcommand{\difftree}{\textsc{Difftree}\xspace}
\newcommand{\difftrees}{{\difftree}s\xspace}

%% file: introduction.tex
\section{Introduction}
SQL is the dominant language for accessing and analyzing data. Along with the recent rise of notebooks, a lot of popular computational notebooks for SQL have emerged such as xeus-sqlite~\cite{sqlkernel} in Jupyter, SQL notebook~\cite{sqlnotebook}, Hex~\cite{hex}, Count~\cite{count},  etc.
Data scientists have gradually ditched their traditional SQL IDEs where they can only see one output at a time in favor of computational notebooks so that they can enjoy narrative programming benefits~\cite{knuth1984literate}. 

Traditional notebooks that can execute SQL queries~\cite{sqlnotebook,sqlkernel} merely render query results as tables.   
This is neither satisfying nor effective, as data scientists rely on interactive visualization interfaces to rapidly perform iterative analyses and to better present analyses in a compelling narrative~\cite{segel2010narrative}.

\begin{table}[]
\small{
\begin{tabular}{rcccc}
  \textit{}                                                                     & \textit{Lux} & \textit{Count} & \textit{Hex} & \textit{\red{\bf PI2}} \\ 
  \textit{Visualizations}                                                                  & \textit{\checkmark}    & \textit{\checkmark}      & \textit{\checkmark}    & \textit{\red{\bf\Checkmark}}    \\ 
  \textit{Widgets}                                                                            & \textit{$\times$}    & \textit{Parameter}      & \textit{Parameter}    & \textit{\red{\bf Arbitrary}}    \\ 
    \textit{Vis.   Interactions}              & \textit{$\times$}    & \textit{$\times$}      & \textit{$\times$}    & \textit{\red{\bf\Checkmark}}    \\ 
      \textit{Zero Effort}                                                                            & \textit{\checkmark}    & \textit{$\times$}      & \textit{$\times$}    & \textit{\red{\bf\Checkmark}}    \\ 

\end{tabular}
}
\caption{Comparison among different tools.}
\vspace{-9mm}
\label{t:compare}
\end{table}

\begin{figure}[t]
  \centering
  \begin{subfigure}[b]{.43\columnwidth}
      \centering
      \includegraphics[width=\columnwidth]{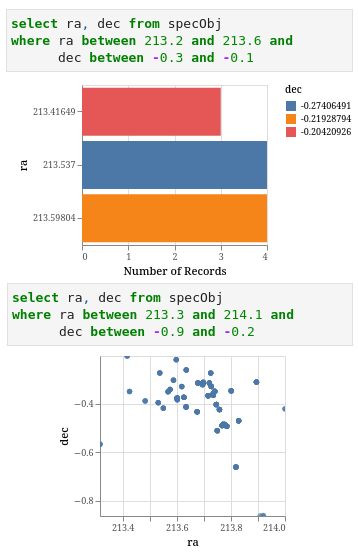}
      \caption{Lux.}
      \label{f:sdss-lux}
    \end{subfigure}
  \begin{subfigure}[b]{.47\columnwidth}
    \centering
      \includegraphics[width=\columnwidth]{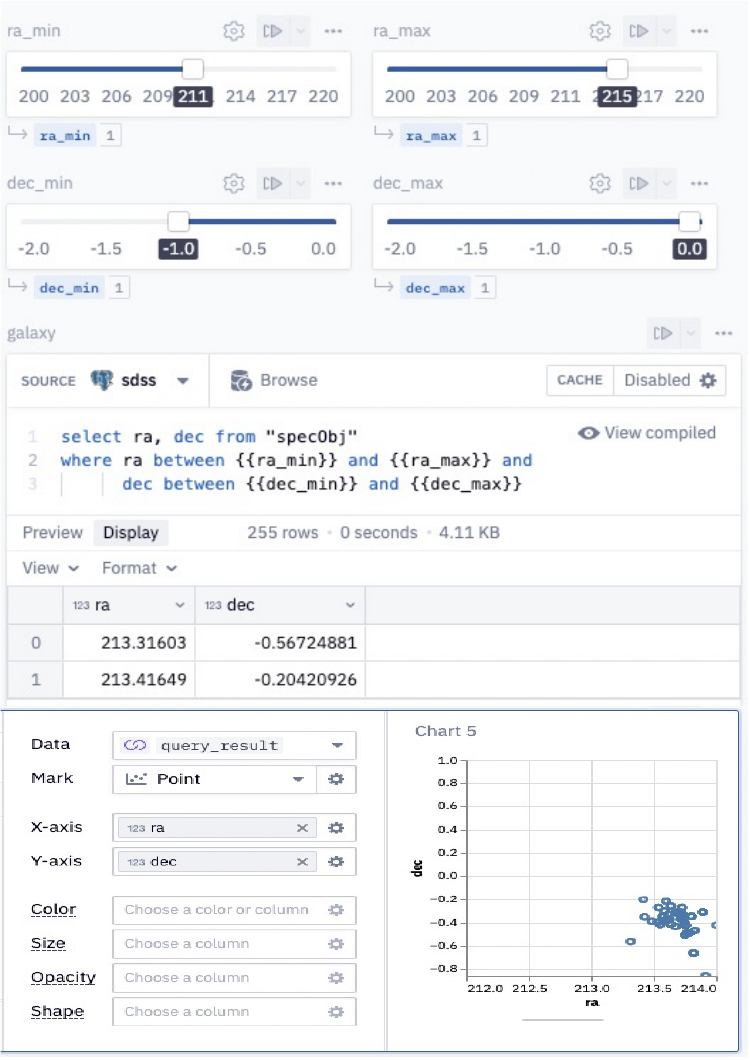}
    \caption{Hex Notebook.}
    \label{f:sdss-hex}
  \end{subfigure}
  \begin{subfigure}[b]{.9\columnwidth}
    \centering
      \includegraphics[width=\columnwidth]{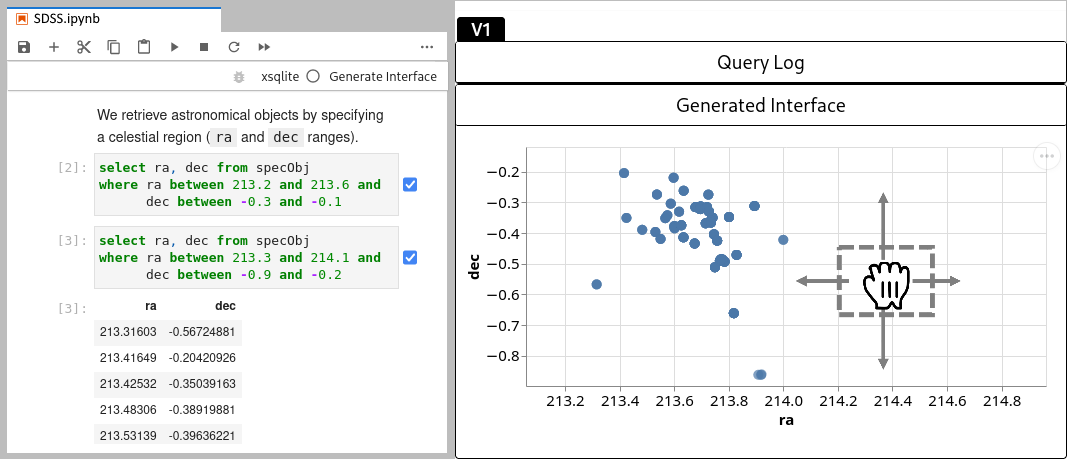}
    \caption{This paper: \sys.}
    \label{f:sdss-pi2}
  \end{subfigure}
  \vspace{-4mm}
  \caption{Different interfaces for analysis of the SDSS dataset: (a) static visualization recommendation with Lux, (b) parameterized query with widgets and visualization with Hex, (c) automatically generated interactive interface with PI2.}
  \vspace{-6mm}
  \label{f:sdss}
\end{figure}

In contrast to static tables, interactive visualization interfaces (or {\it interfaces}) consist of three main components: visualizations (e.g., bar and line charts), widgets (e.g. dropdown, slider), and interactions within a visualization (e.g. brushing to select points, panning, clicking).
As such, numerous recent notebooks and extensions, such as Lux~\cite{lee2021lux}, Count Notebook~\cite{count}, and Hex Notebook~\cite{hex} have been designed to help users visualize data and create simple interactive visualizations during their analysis.
Unfortunately, the type and complexity of interfaces that they can express are limited (\Cref{t:compare}).  
For instance, Lux~\cite{lee2021lux} automatically recommends a static visualization when a notebook cell returns a dataframe, but does not support interactive analysis. 
Similarly, Count~\cite{count} and Hex~\cite{hex} let users visualize a table and add custom widgets that manipulate simple query parameters, but these require explicit user effort.
In short, existing notebooks have limited support for interactivity, do not support generating interactive visualizations, and require manual effort to create and lay out visualizations and widgets.

This paper demonstrates \sys, the first notebook extension which can automatically generate interactive visualization interfaces during SQL-based analyses. Users can select relevant queries during their analysis and invoke \sys to synthesize a fully interactive interface with no additional effort.    \sys automatically chooses the appropriate visualizations, widgets, and visualization interactions to fully express the analysis represented by the user's selected queries.

\begin{example}
  \Cref{f:sdss} illustrates the types of interfaces that Lux, Hex, and \sys will generate using queries from the Sloan Digital Sky Survey(SDSS)~\cite{sdss} query log.   The two example queries retrieve astronomical objects by specifying a celestial region ({\tt ra} and {\tt dec} ranges). Lux recommends visualizations for individual tables, so it generates different visualizations for each query, despite their similarities (\Cref{f:sdss-lux}).  Note that the users will need to repeatedly tweak and re-execute the queries if they continue their analysis.
 
  Hex lets the user parameterize the {\tt ra} and {\tt dec} values in the query, create custom sliders to control each parameter, and visualize the query results in a visualization (\Cref{f:sdss-hex}). This enables more interactivity than Lux, since the user can directly use the sliders instead of editing SQL strings.  However, the user still needs to configure the sliders and choose an effective visualization.  Furthermore,  {\it using} the interface is cumbersome, as the user needs to manipulate four separate sliders to pan and zoom.
  
In contrast, \sys uses the same two queries to generate the interface in \Cref{f:sdss-pi2}.   The visualization supports panning and zooming, so the user can simply drag and scroll on the visualization to manipulate the {\tt ra} and {\tt dec} ranges and receive immediate visual feedback. The collapsed Query Log tab archives the input queries.
\end{example}

In addition to support for visualization interactions, \sys goes beyond existing notebooks in two ways.   First, the widgets and interactions are more than simple query parameters, and can change the structure of the underlying SQL queries as well.  For instance, a dropdown may choose between three subqueries, a switch may toggle a filter clause, and a tab may select between different queries to visualize.      Second, \sys takes the available screen size into account in order to select a good layout for the interface---on a large screen, the interface may show multiple visualizations side by side, whereas a small screen may show a single visualization that can be changed via interactions. 

We further design the notebook extension to aid iterative analyses.  The atomic unit of execution in notebooks allows users to iteratively refer back to previous cells to edit and potentially re-execute them. To adapt to edits and ensure the reproducibility of the generated interface, we take a snapshot of the queries used to generate a new interface. We also version the interfaces, so that users can go back to, or fully revert, to a previous analysis. To avoid interruption of the normal notebook workflow, we choose to lay the {\it Generated Interface} panel side-by-side with the notebook cells.

The next section presents a brief overview of how \sys generates interfaces from queries, and  \Cref{s:demo} will illustrate these features in the context of a case study.  We refer readers to the technical report~\cite{chen2021pi2} for complete technical details of the interface generation process.

%% file: overview.tex
\section{Interface Generation Overview}\label{s:overview}

\begin{figure}
  \centering
  \includegraphics[width=\columnwidth]{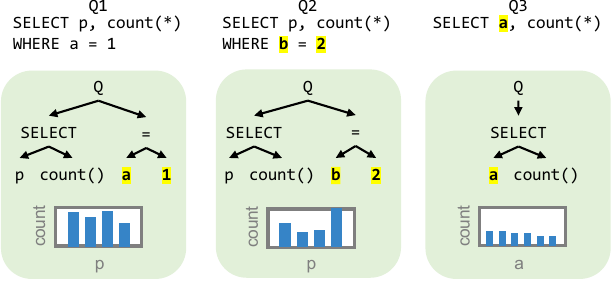}
    \vspace{-7mm}
  \caption{Example of three queries and their simplified ASTs.  A static interface would render one chart for each query.}
    \vspace{-3mm}
  \label{f:log}
\end{figure}

\begin{figure}
  \centering
  \includegraphics[width=0.7\columnwidth]{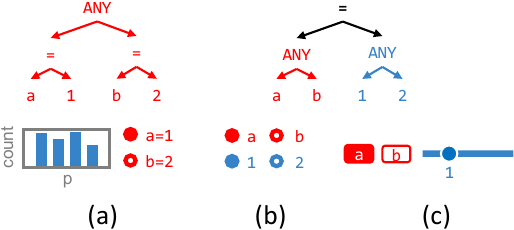}
  \vspace{-3mm}
  \caption{ Example \difftrees and interfaces for \texttt{Q1}, \texttt{Q2}, focusing on the subtree for the predicate.  The \texttt{ANY} choice node can choose one of its children.  (a) the \texttt{ANY} node maps to a radio button that chooses between the two predicates, 
    (b) the radio lists separately specify the left and right operands, (c) the choice nodes can instead be mapped to a button group and slider, and organized vertically.    }
  \vspace{-3mm}
  \label{f:q12}
\end{figure}

\sys transforms an input sequence of queries into an interactive interface in four steps: 
it parses queries into a generalization of abstract syntax trees (ASTs) that we call \difftrees; maps the \difftrees to a candidate interface; estimates the interface's cost; and repeatedly transforms the \difftrees to generate new candidate interfaces, optimizing according to cost. This section walks through these steps and introduces key concepts.

\stitle{Static Interfaces: }
\Cref{f:log} lists three queries\footnote{For brevity, we omit the \texttt{FROM} and \texttt{GROUPBY} clauses and show simplified ASTs.} and their ASTs, where attributes \texttt{p}, \texttt{a}, \texttt{b} are integers.  \texttt{Q1} and \texttt{Q2} change the predicate attribute and literal, and \texttt{Q3} selects \texttt{a} instead of \texttt{p}.  Since each AST is a \difftree, a valid interface simply maps each AST's results to a static chart.

\stitle{Interactive Interfaces:}
Let us focus on the differing predicate in \texttt{Q1} and \texttt{Q2} to show how
different \difftree structures can result in different interface designs.
For instance, \Cref{f:q12}(a) is rooted at an \texttt{ANY} node whose children are the two predicates.  \texttt{ANY} is a {\it Choice Node} that can choose one of its child subtrees.
In general, choice nodes encode subtree variations\footnote{Choice nodes generalize SQL parameterized literals to syntactic structures. } that the user can control through the interface.
In the example, the \texttt{ANY} node is mapped to two radio buttons (other widgets such as a dropdown are valid as well), where clicking on the first
button would bind the \texttt{ANY} to its first child \texttt{a=1}.  
The \difftree output is visualized as a bar chart.

\ititle{Tree Transformations:} Both of \texttt{ANY}'s children are rooted at \texttt{=}, so the \texttt{=} can be refactored above the \texttt{ANY} node.  This is an example of a {\it Tree Transformation Rule}.
The resulting \difftree in \Cref{f:q12}(b) shows two \texttt{ANY}
nodes that can independently choose the left and right operands.  This leads 
to an interface with two interactions---two sets of radio buttons---and also generalizes the interface beyond
the input queries.  For instance, the query can now express \texttt{SELECT p, count(*) WHERE b=1}.

A \difftree can map to many interface designs, each with different visualizations, interactions (including widgets and visualization interactions), and layouts.  For instance, \Cref{f:q12}(b) and (c) both express \texttt{Q1} and \texttt{Q2}, however \Cref{f:q12}(c) uses horizontal layout and the slider can choose a continuous range of numbers that generalize beyond the radio buttons in \Cref{f:q12}(b).

\begin{figure}
  \centering
  \includegraphics[width=.8\columnwidth]{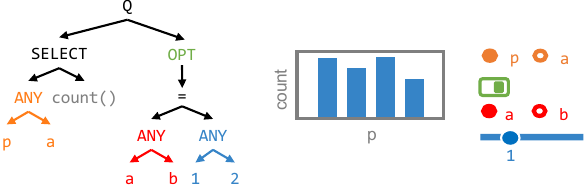}
    \vspace{-2mm}
  \caption{ A \difftree for \texttt{Q1-3} and a candidate interface}
  \vspace{-2mm}
  \label{f:q123}
\end{figure}

\ititle{All Three Queries: } Now, let us add \texttt{Q3}.   
A simple approach would be to partition the queries into two clusters, 
where  \texttt{Q3} is rendered as a static chart, 
and \texttt{Q1} and \texttt{Q2} are mapped to one of the interfaces discussed so far. 
We can then choose to lay these two visualizations out horizontally or vertically.
Another possibility is to merge all three queries into a single \difftree,
which would map to an interface with a single visualization.
\Cref{f:q123} illustrates one such \difftree structure, 
where an \texttt{ANY} node in the \texttt{SELECT} clause chooses whether to project \texttt{p} or \texttt{a}.
This maps to an interface similar to \Cref{f:q12}(c), but with a radio button to choose the attribute to project and a toggle for the optional \texttt{WHERE} clause.
Naturally, which of these possible interface designs (or others not discussed here)
that should be generated and returned to the user depends on many factors, 
such as usability, layout, accessibility, and other factors that are difficult to quantify.  
Quantitative interface evaluation is an active area of research, and 
\sys borrows current best practices to develop its cost function.

\begin{figure}
  \centering
  \includegraphics[width=\columnwidth]{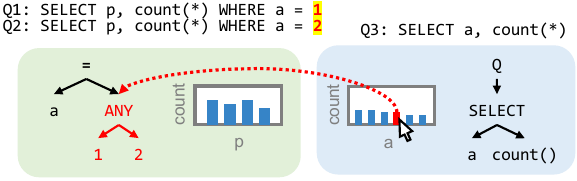}
    \vspace{-8mm}
  \caption{ Multi-view interface where clicking on the right-side chart updates the left chart.}
    \vspace{-6mm}
  \label{f:multiview}
\end{figure}

\stitle{Multi-view Interfaces}
\sys can also generate  multi-view interfaces.   
\Cref{f:multiview} illustrates a slightly different set of queries,
where the \texttt{Q1} and \texttt{Q2} only differ in the literal, and \texttt{Q3} remains the same.
Since the literal is compared to attribute \texttt{a}, an alternative to mapping 
the \texttt{ANY} node to a slider is to map it to a {\it visualization interaction}
in \texttt{Q3}'s bar chart. Specifically, each bar is derived from \texttt{(a, count(*))}
in \texttt{Q3}'s result. Thus, clicking on a bar can also derive a valid
value in attribute \texttt{a}'s domain that can bind to the \texttt{ANY} node.

\stitle{Summary and Generation Pipeline:}

\begin{figure}[t]
  \centering
  \includegraphics[width=\columnwidth]{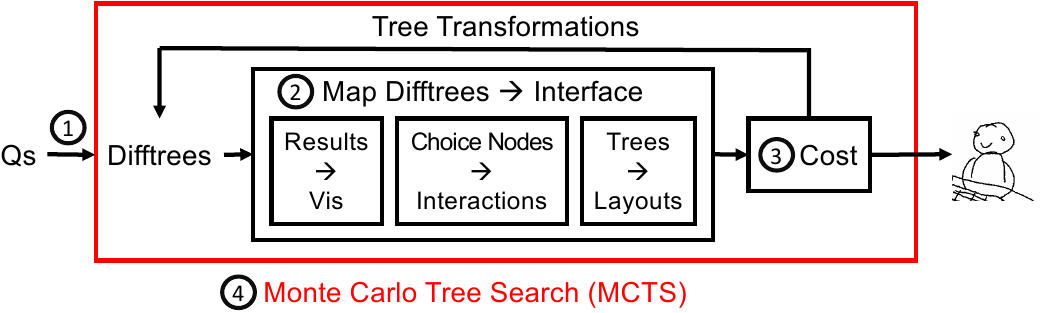}
  \vspace{-8mm}
  \caption{\sys interface generation pipeline.}
    \vspace{-6mm}
  \label{f:pipeline}
\end{figure}

\sys transforms an input sequence of queries into an interactive interface in four steps (\Cref{f:pipeline}).
\textcircled{1} It first parses the input query sequence $Q$ into \difftrees. 
\textcircled{2} \sys maps the \difftrees
to an interface.  
An interface mapping $\mathbb{I}=(\mathbb{V},\mathbb{M},\mathbb{L})$ is defined by a \textit{Visualization Mapping} $\mathbb{V}$ from \difftrees results to visualizations, a \textit{Interaction Mapping} $\mathbb{M}$ from choice nodes to interactions (including widgets and visualization interactions),
and a \textit{Layout Mapping} $\mathbb{L}$ from \difftrees structures to layouts. We formulate the interface mapping problem as a schema matching problem by defining schema for both \difftrees and interfaces.
\textcircled{3} A cost model $C(\mathbb{I}, Q)$ evaluates the interface and \sys either returns the interface or 
chooses a valid transformation to apply to the \difftrees.  
Informally, our problem is to return the lowest cost interface $\mathbb{I}$ that can express all queries in $Q$.
\textcircled{4} The space of possible interfaces is enormous, so we solve this problem using Monte Carlo Tree Search~\cite{Browne2012ASO} (MCTS).
MCTS balances exploitation of good explored states (\difftree structures) with exploration of new states.

%% file: demontration.tex
\begin{figure*}[h]
    \centering
    \includegraphics[width=\textwidth]{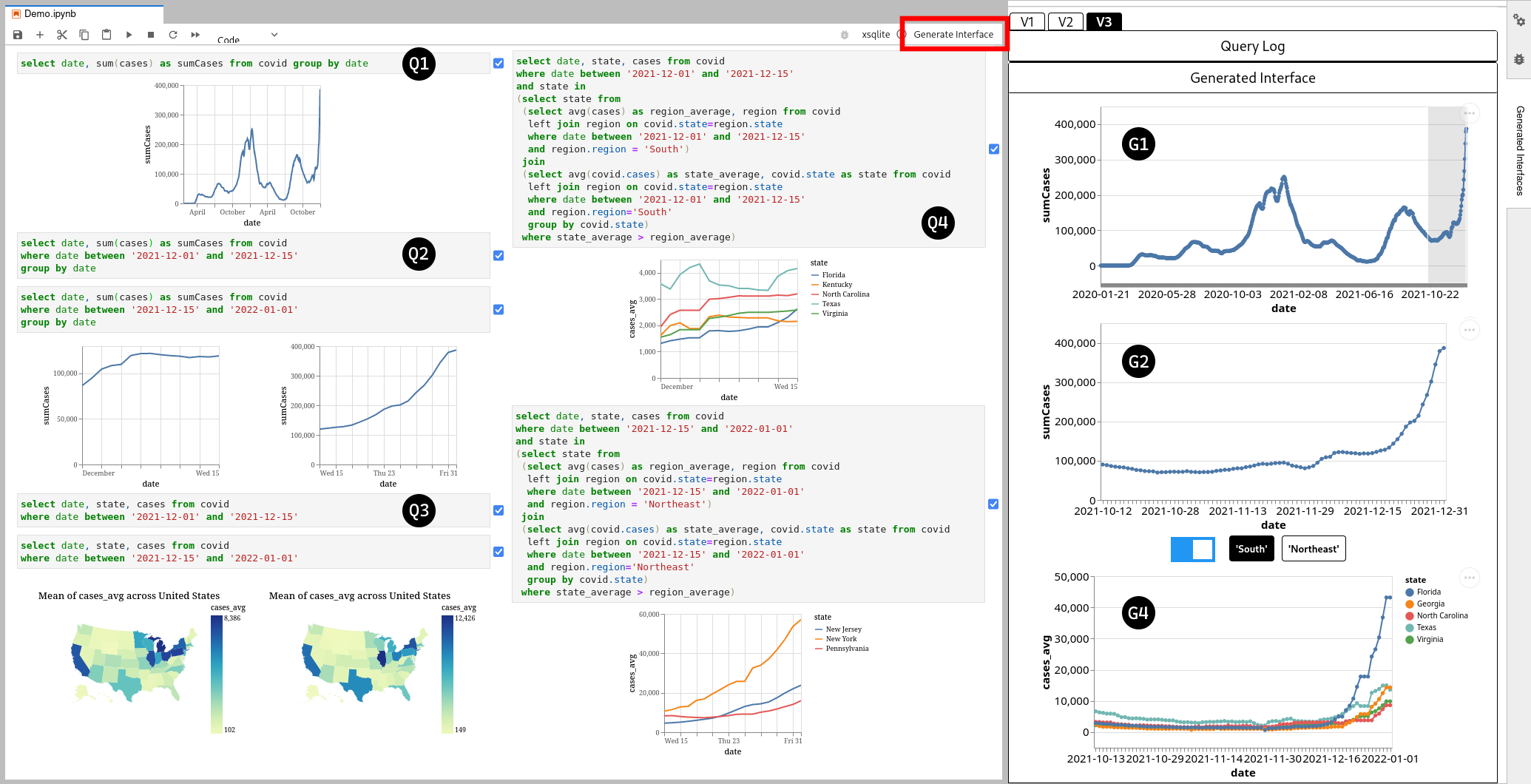}
    \vspace{-5mm}
    \caption{\sys Jupyter extension interface. Left: SQL-based analyses. Right: An interface automatically generated by \sys.}
    \label{fig:demo}
    \vspace{-5mm}
\end{figure*}

\section{Demonstration}
\label{s:demo}    
\subsection{Interface Design}
 We integrate \sys as a Jupyter Lab~\cite{granger2016jupyterlab} extension. 
 We design the interface as shown in \Cref{fig:demo}. 
 While authoring SQL queries in the Jupyter notebook, a user can check the checkbox next to each cell to include it as part of the query log for interface generation. 
 Clicking the {\small {\tt Generate Interface}} button will invoke \sys to generate a new interface to the \textit{Generated Interfaces} panel on the right.
 
The atomic unit of execution in notebooks allows users to easily refer back to previous cells to edit and potentially re-execute them. To adapt to edits and ensure reproducibility, our integration tracks interface versions in the version tabs at the top of the \textit{Generated Interfaces} panel and archives the input query logs in the \textit{Query Log} collapsible section for each version. \sys lays out the interfaces and notebook cells side-by-side, so that the normal notebook analysis workflow will not be interrupted.

\subsection{Use Case Walkthrough}
We demonstrate how \sys aids the data analysis process via a real-world scenario (shown in \Cref{fig:demo}): in late December 2021, an analyst named Jane at a news organization is analyzing a COVID-19 dataset of daily case counts per-state with the intent to give travel warning advice for the winter holiday season. 
For the sake of comparison, we show a static visualization recommendation below each cell, which is given by an existing system Lux~\cite{lee2021lux}.

\stitle{Step 1: Overview and detailed look of the dataset.}
Seeking to get an overall view of the data, Jane writes {\tt Q1} and gets a visualization recommended by Lux showing total case count over time. Looking for a more detailed view, Jane restricts the date range in {\tt Q2} to look back over two preceding half-month periods to see more recent trends. Moreover, she would like to do this over different date ranges which will result in many similar static visualizations and a lengthy notebook. With \sys, she can select these three queries via their corresponding checkboxes and automatically generate an interface.
\sys produces a unified interactive interface V1 consisting of two plots: one showing the overall timeline (G1) and the other showing just the selected date range (G2). The two plots are linked by a brushing interaction so that brushing over G1 dynamically configures the date range of G2. Whereas, none of existing notebook tools can create such visualization interactions. In this way, Jane can use this interface to further investigate trends in different date ranges interactively and without writing more SQL queries.

\stitle{Step 2: Drill down into state Level.}
 Jane's task is to give travel warning advice. She writes {\tt Q3} to study each state's trends over time. Lux generates choropleth maps visualizing the case count for each state averaged over the date window, which do not allow her to see trends over time. Alternatively, she can use \sys to generate a new interface V2 with three plots: two are the sames as interface V1 and the third chart (G3) is a bar chart $(x\rightarrow date, y\rightarrow cases, color\rightarrow state)$. We omit G3 in \Cref{fig:demo} due to space constraints. In this interface V2, the linked brushing interaction will configure the queries underlying both G2 and G3 such that Jane can brush over G1 to see the detailed trend and per-state breakdown trend within the selected date range at the same time.

\stitle{Step 3: Focused region investigation.}
However, the state breakdown view is visually noisy due to the large number of states. Jane decides to group states into regions and only show those states whose average case counts over a period of time exceed the region's average, expressed as the complicated query {\tt Q4} consisting of joins, and correlative subqueries. Further, Jane would like to study the South and Northeast regions within different date ranges. She selects all the queries and invokes \sys. 
The interface V3 that \sys generates in \Cref{fig:demo} has three plots, allowing Jane to view the overall timeline (G1), detail view of the selected date range (G2), and state breakdown filtered for above average states (G4). This new interface maintains the date brushing functionality of previous versions and introduces query configuration widgets: a toggle which allows her to toggle between G3 and G4, and a pair of buttons that switch between the South and Northeast regions. Structurally, the toggle corresponds to an {\tt OPT} choice node that distinguishes the existence of a complicated subquery---{\small{\tt \{and state in...\}}} in Q4 not present in Q3. 
Through this interface, Jane is able to fluidly reconfigure the date range by brushing on G1, toggle off to see the overall state breakdown of cases, and toggle on to choose to observe trends in the Northeast or South. Noticing very high rates of growth in case count, Jane makes a recommendation that travelers avoid Florida in the South and New York in the Northeast.

Through the above scenario, we show \sys's ability to consume arbitrarily complex SQL queries and automatically generate complete interfaces, including visualization interactions and widgets expressing arbitrary query differences that no other tools can.

\stitle{Demonstration engagement.} Participants will be able to write their analysis and generate interactive visual interfaces using \sys Jupyter extension. We will prepare three datasets – \texttt{COVID-19}, \texttt{SDSS}, and \texttt{S\&P500} for users to explore.